\documentclass[twocolumn,prb,final,amsmath,amssymb,superscriptaddress,showpacs]{revtex4-2}
\usepackage{amsmath}
\usepackage{amssymb}
\usepackage[colorlinks=true,citecolor=blue]{hyperref}
\usepackage{graphicx}
\usepackage{babel}

\makeatletter
\usepackage{graphics}\usepackage{color}

\usepackage{epsfig}
\usepackage{bm}
\usepackage{bbm}

\definecolor{Red}{rgb}{1,0,0}
\definecolor{Blu}{rgb}{0,0,1}
\definecolor{Green}{rgb}{0,1,0}

\makeatother
\usepackage{amssymb}
\usepackage{babel}

\usepackage{tikz,xcolor,hyperref}
\definecolor{lime}{HTML}{A6CE39}
\DeclareRobustCommand{\orcidicon}{%
	\begin{tikzpicture}
	\draw[lime, fill=lime] (0,0)
	circle [radius=0.16]
	node[white] {{\fontfamily{qag}\selectfont \tiny ID}};
	\draw[white, fill=white] (-0.0625,0.095)
	circle [radius=0.007];
	\end{tikzpicture}
	\hspace{-2mm}
}

\foreach \x in {A, ..., Z}{%
	\expandafter\xdef\csname orcid\x\endcsname{\noexpand\href{https://orcid.org/\csname orcidauthor\x\endcsname}{\noexpand\orcidicon}}
}

\usepackage{dsfont}
\usepackage{mathdots}
\usepackage{amsfonts}
\usepackage{latexsym}
\usepackage[normalem]{ulem}
\usepackage{mathtools}

\makeatother

\begin{document}
\title{Unprotected edge modes in quantum spin Hall insulator candidate materials}

\author{Nguyen Minh Nguyen\orcidF}
\affiliation{International Research Centre MagTop, Institute of Physics, Polish Academy of Sciences, Aleja Lotnik\'ow 32/46, PL-02668 Warsaw, Poland}

\author{Giuseppe Cuono\orcidA}
\affiliation{International Research Centre MagTop, Institute of Physics, Polish Academy of Sciences, Aleja Lotnik\'ow 32/46, PL-02668 Warsaw, Poland}

\author{Rajibul Islam\orcidB}
\affiliation{International Research Centre MagTop, Institute of Physics, Polish Academy of Sciences, Aleja Lotnik\'ow 32/46, PL-02668 Warsaw, Poland}

\author{Carmine Autieri\orcidC}
\email{autieri@magtop.ifpan.edu.pl}
\affiliation{International Research Centre MagTop, Institute of Physics, Polish Academy of Sciences, Aleja Lotnik\'ow 32/46, PL-02668 Warsaw, Poland}
\affiliation{Consiglio Nazionale delle Ricerche CNR-SPIN, UOS Salerno, I-84084 Fisciano (Salerno),
Italy}

\author{Timo Hyart\orcidD}
\email{timo.hyart@tuni.fi}
\affiliation{Computational Physics Laboratory, Physics Unit, Faculty of Engineering and Natural Sciences, Tampere University, FI-33014 Tampere, Finland}
\affiliation{Department of Applied Physics, Aalto University, 00076 Aalto, Espoo, Finland}
\affiliation{International Research Centre MagTop, Institute of Physics, Polish Academy of Sciences, Aleja Lotnik\'ow 32/46, PL-02668 Warsaw, Poland}

\author{Wojciech Brzezicki\orcidE}
\email{brzezicki@magtop.ifpan.edu.pl}
\affiliation{International Research Centre MagTop, Institute of Physics, Polish Academy of Sciences, Aleja Lotnik\'ow 32/46, PL-02668 Warsaw, Poland}
\affiliation{Institute of Theoretical Physics, Jagiellonian University, ulica S. \L ojasiewicza
11, PL-30348 Krak\'ow, Poland}

\date{\today}
\begin{abstract}
The experiments in quantum spin Hall insulator candidate materials, 
such as HgTe/CdTe and InAs/GaSb heterostructures, indicate that in 
addition to the topologically protected helical edge modes these 
multilayer heterostructures may also support additional  edge 
states, which can contribute to the scattering and the transport. We use first-principles calculations to derive an effective 
tight-binding model for HgTe/CdTe, HgS/CdTe and InAs/GaSb heterostructures, and we show that all these materials support additional edge states which are sensitive to the edge termination. We trace the microscopic origin of these states back to a minimal model supporting flat bands with a nontrivial quantum geometry that gives rise to polarization charges at the edges. We show that the polarization charges transform into the additional edge states  when the flat bands are coupled to each other and to the other states to form the Hamiltonian describing the full heterostructure. Interestingly, in the HgTe/CdTe quantum wells the additional edge states  are far away from the 
Fermi level so that they do not contribute to the transport but  in the HgS/CdTe and InAs/GaSb heterostructures they appear within the bulk energy gap giving rise to the possibility of multimode edge transport. Finally, we demonstrate that because these additional edge modes are non-topological it is possible to remove them from the bulk energy gap  by modifying the edge potential for example with the help of a side gate or chemical doping.
\end{abstract}
\maketitle

\section{Introduction}

The theory of quantum spin Hall (QSH) effect predicts the existence 
of helical edge modes, which are topologically protected against 
elastic backscattering from all perturbations obeying the time-reversal symmetry, 
and various materials have been predicted to support the QSH insulator phase 
\cite{RevKane, RevZhang, BHZ, Liu08}. Experimentally, signatures of edge mode 
transport have been observed in several of the candidate materials such as 
HgTe/CdTe quantum wells \cite{QSH_HgTe}, InAs/GaSb bilayers \cite{Du15} and 
WTe$_2$ \cite{Wu18}. However, the experimental studies have also led to 
discrepancies with the simple theoretical models. In WTe$_2$ the protection length of the edge transport is only few tens of nanometers \cite{Wu18}, and even in the 
more extensively studied InAs/GaSb and HgTe/CdTe quantum wells the best protection lengths reached so far are on the order of few micrometers \cite{Du15} and few tens of micrometers \cite{PhysRevLett.123.047701}, respectively. There is still no consensus about the interpretation of the observed short protection lengths but various mechanisms, such as magnetic impurities \cite{PhysRevLett.106.236402}, phonons \cite{PhysRevLett.108.086602}, dynamic nuclear polarization \cite{PhysRevB.86.035112,PhysRevB.87.165440}, spontaneous time-reversal symmetry breaking \cite{Pikulin14, Paul1}, charge puddles \cite{Vayrynen14}, charge dopants \cite{Dietl2022} and interaction effects \cite{Dolcetto16}, may contribute to the breakdown of the topological protection.

The quality of the edge  transport can be improved with the help of impurity doping in InAs/GaSb bilayers \cite{Du15} and gate training in HgTe/CdTe quantum wells \cite{PhysRevLett.123.047701} indicating that there likely exists some additional unprotected low-energy states, which are contributing to the breakdown of the topological protection and which are influenced by these sample preparation techniques. In certain experiments unprotected edge states have been observed also more directly. Namely, in InAs/GaSb bilayers in the absence of the impurity doping multimode edge transport has been experimentally observed in the trivial regime \cite{Nichele_2016}. In HgTe/CdTe quantum wells the additional states seem to be sufficiently far away from the Fermi level so that they do not contribute to the transport, but the dynamical properties suggest that the topological edge states are surrounded by additional states contributing  to the scattering \cite{PhysRevLett.124.076802}. The microscopic origin of these additional states remain unknown in both materials.

In this paper, we use first-principles calculations to derive an effective tight-binding model for HgTe/CdTe, HgS/CdTe and InAs/GaSb heterostructures, and we show that all these materials support additional edge states which are sensitive to the edge termination. We trace the microscopic origin of these states back to a minimal model of a buckled honeycomb lattice of anions and cations. This system is the minimal building block for constructing the HgTe/CdTe, HgS/CdTe and InAs/GaSb heterostructures, and it supports flat bands with nontrivial quantum geometry that gives rise to polarization charges at the edges \cite{Rhi17}. We show that the polarization charges transform into the additional edge states  when the flat bands are coupled to each other and to the other states to form the Hamiltonian describing the full heterostructure. In the HgTe/CdTe quantum wells the additional edge states  are far away from the 
Fermi level so that they do not contribute to the transport but  in the HgS/CdTe and InAs/GaSb heterostructures they appear within the bulk energy gap giving rise to the possibility of multimode edge transport in agreement with the experiments \cite{Nichele_2016, PhysRevLett.124.076802}. Finally, we demonstrate that because these additional edge modes are non-topological it is possible to remove them from the bulk energy gap  by modifying the edge potential for example with the help of a side gate or chemical doping, providing a possible explanation for the mysterious improvement of the quality of the QSH effect with the help of impurity doping in InAs/GaSb bilayers \cite{Du15} and gate training in HgTe/CdTe quantum wells \cite{PhysRevLett.123.047701}.

\section{Heterostructure Hamiltonian and non-topological edge states}

\begin{figure}[b]
\includegraphics[width=0.6\columnwidth]{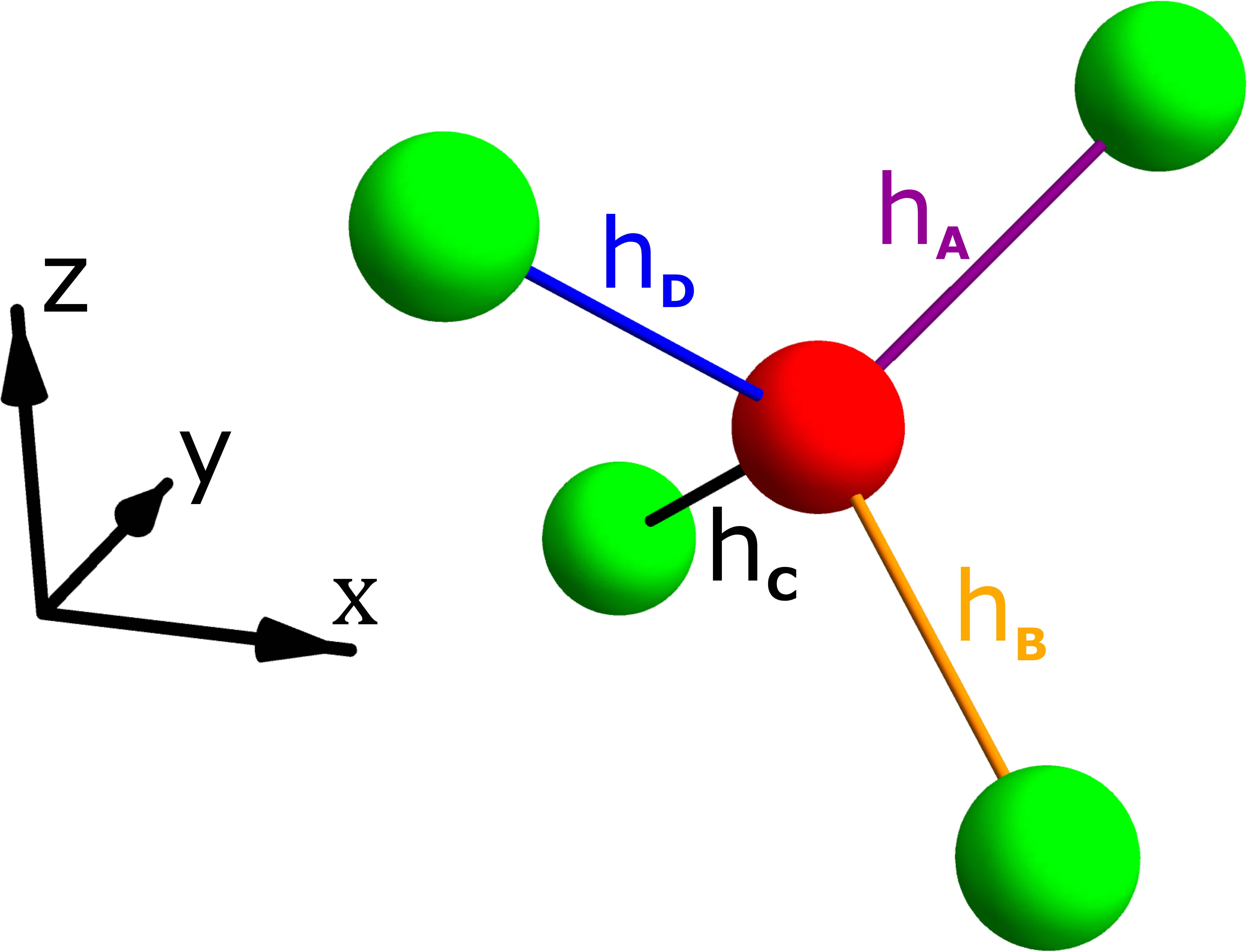}\caption{Schematic illustration of the hopping matrices $h_\alpha$ ($\alpha=A, B, C, D$) between the nearest-neighbour lattice sites in zinc-blende crystals. Each  cation (green) and anion (red) supports one $s$-orbital and three $p$-orbitals so that $h_i$ are 4x4 matrices. \label{fig:1}}
\end{figure}

Our starting point are the Hamiltonians for HgTe, HgS, CdTe, InAs, GaSb and AlSb bulk crystals
\begin{widetext}
\begin{eqnarray}
\mathcal{H}(\textbf{k}) &=&   \mathds{1}_2\!\otimes\!h_A\!\otimes\!\begin{pmatrix}
0 & e^{ik_{3}} \\
 0&0 
\end{pmatrix} + \mathds{1}_2\!\otimes\!h_B\!\otimes\!\begin{pmatrix}
0 & e^{ik_{1}} \\
 0&0 
\end{pmatrix} +\mathds{1}_2\!\otimes\!h_C\!\otimes\!\begin{pmatrix}
0 & 1 \\
 0&0 
\end{pmatrix} + \mathds{1}_2\!\otimes\!h_D\!\otimes\!\begin{pmatrix}
0 & e^{i\left( k_{3} -k_{2}\right)}\\
 0&0 
\end{pmatrix} + \mathit{h.c}\nonumber \\
&&\hspace{-0.8cm}+   \mathds{1}_2 \!\otimes\! \mathit{E}_{c} \!\otimes\! \begin{pmatrix}
1 & 0\\
 0&0 
\end{pmatrix} +   \mathds{1}_2 \!\otimes\! \mathit{E}_{a} \!\otimes\! \begin{pmatrix}
0 & 0\\
 0&1 
\end{pmatrix} -   \sum_{\alpha=1,2,3} \!\!   \sigma_{\alpha} \!\otimes\! \mathit{L}_{\alpha} \otimes \begin{pmatrix}
\lambda_c & 0\\
 0& \lambda_a 
\end{pmatrix}    - \mathit{E}_{\mathit{f}} \!\otimes\! \mathds{1}_{16},
\label{eq:Ham3D}
\end{eqnarray}
\begin{equation}
 h_A=  h\left(\dfrac{1}{\sqrt{3}},\dfrac{1}{\sqrt{3}},\dfrac{1}{\sqrt{3}} \right),
 h_B=  h\left(\dfrac{1}{\sqrt{3}},-\dfrac{1}{\sqrt{3}},-\dfrac{1}{\sqrt{3}} \right),
 h_C=  h\left(-\dfrac{1}{\sqrt{3}},\dfrac{1}{\sqrt{3}},-\dfrac{1}{\sqrt{3}} \right),
h_D=  h\left(-\dfrac{1}{\sqrt{3}},-\dfrac{1}{\sqrt{3}},\dfrac{1}{\sqrt{3}} \right)
\end{equation}
\begin{equation}
h\left(l,m,n \right) = 
 %\begin{small}
\begin{pmatrix}
V_{ss\sigma} & lV_{s_c p_a\sigma} & mV_{s_c p_a\sigma}& nV_{s_c p_a\sigma}  \\
 -lV_{s_a p_c\sigma} &  l^2 \left(V_{p p\sigma}-V_{p p\pi} \right)+V_{p p\pi}&  lm \left(-V_{p p\pi}+V_{p p\sigma} \right)&  ln\left(-V_{p p\pi}+V_{p p\sigma} \right) \\
   -mV_{s_a p_c\sigma} &  lm \left(-V_{p p\pi}+V_{p p\sigma} \right)  & m^2 \left(V_{p p\sigma}-V_{p p\pi} \right)+V_{p p\pi}  & mn \left(-V_{p p\pi}+V_{p p\sigma} \right)\\
   -nV_{s_a p_c\sigma}&  ln \left(-V_{p p\pi}+V_{p p\sigma} \right)&  mn \left(-V_{p p\pi}+V_{p p\sigma} \right)&  n^2 \left(V_{p p\sigma}-V_{p p\pi} \right)+V_{p p\pi} \label{eq:hop} 
\end{pmatrix}
%\end{small}
\end{equation}
\begin{equation}
L_x= i \begin{small}\begin{pmatrix}
0 & 0& 0&0 \\
 0 & 0& 1&0 \\
 0 & -1& 0&0 \\
 0 & 0& 0&0 
\end{pmatrix},\end{small} \ 
 L_y= i \begin{small}\begin{pmatrix}
0 & 0& 0&0 \\
 0 & 0& 0&-1 \\
 0 & 0& 0&0 \\
 0 & 1& 0&0 \end{pmatrix},\end{small} \ 
  L_z= i \begin{small}\begin{pmatrix}
0 & 0& 0&0 \\
 0 & 0& 0&0 \\
 0 & 0& 0&1 \\
 0 & 0& -1&0 
\end{pmatrix} \end{small}, \ E_{a(c)}= \begin{small} \begin{pmatrix}
V_{s_{a(c)}} & 0& 0&0 \\
 0 & V_{p_{a(c)}}& 0&0 \\
 0 & 0& V_{p_{a(c)}}&0 \\
 0 & 0& 0&V_{p_{a(c)}} 
\end{pmatrix},\end{small}
\end{equation}
\end{widetext}
where the tight-binding parameters for each material, given in 
Table~\ref{table:1}, have been derived from first-principles 
density-functional theory (DFT) calculations. 
To obtain the tight-binding with first-neighbour hopping parameters, we impose to the tight-binding model to fit the DFT band structure at high-symmetry points extracting the  on-site energies, the hopping amplitudes and the spin-orbit couplings as fitting parameters.
More technical information are provided in the Appendix \ref{app:DFT}.
Here, the unit cell of the zinc-blende crystal structure contains two lattice sites, anions and cations, at positions $(0,0,0)$ and $(1/2,1/2,1/2)$ with lattice translation vectors being $\textbf{n}_1=(1,1,0)$, $\textbf{n}_2=(-1,1,0)$ and $\textbf{n}_3=(0,1,1)$.
Each site in the unit cell supports one $s$-orbital and three $p$-orbitals, $h_\alpha$ are $4 \times 4$ matrices
describing the hopping amplitudes  between the different orbitals of cations and anions [parametrized by $V_{ss\sigma}$, $V_{s_a p_c \sigma}$, $V_{s_c p_a \sigma}$, $V_{pp\sigma}$ and $V_{pp\pi}$ in Eq.~(\ref{eq:hop})] along the different directions $\alpha=A,B,C,D$ as depicted in Fig.~\ref{fig:1}, matrices $E_{a(c)}$ contain the on-site energies $V_{s_{a(c)}}$ and $V_{p_{a(c)}}$ of the $s$- and $p$-orbitals of the anions (cations), $\sigma_{\alpha}$ are Pauli spin matrices, $L_{\alpha}$ are the $4 \times 4$ angular momentum  matrices,  $\lambda_{a(c)}$ is the spin-orbit coupling strength at the anion (cation) site, and $E_f$ is the Fermi energy.  

\begin{widetext}
\begin{center}
\begin{table}
\begin{tabular}{ |p{1.6 cm}||p{1.5cm}|p{1.5cm}|p{1.5cm}|p{1.5cm}|p{1.5cm}|p{1.5cm}|  }
% \hline
% \multicolumn{4}{|c|}{Parameter table} \\
 \hline
 Parameters& HgTe &HgS&CdTe & InAs & GaSb &AlSb\\
 \hline
 $V_{s_a}$  & -5.8329  & -12.1315   &-6.1832 &   -7.8000  & -5.4804   &-4.2606\\
 $V_{s_c}$&    0.2069    &-1.5535&   1.6395 & -3.5834   &-5.5334&  -3.0847\\
 $V_{p_a}$ &3.1483 & -1.1909&  2.3251 &-0.1424 & -0.2514&  0.4210\\
 $V_{p_c}$     &7.6916 & 5.4898&  7.4584 & 4.1314 & 2.8382&  3.9110\\
 $V_{ss\sigma}$&   -1.2569  & -0.1162&-1.2431  &-0.1424 & -0.2514&  0.4210\\
$V_{s_a p_c\sigma}$ & 1.7229  & 2.8306   &1.6379 &  -1.4257   & -1.5325 &-1.6150\\
 $V_{s_c p_a\sigma}$& 1.4834  & 1.1517&1.5463& 1.4669  & 1.2761& 1.3486\\
 $V_{pp\sigma}$&   2.2132  & 1.5759&2.0139&   2.2223  & 2.200 & 2.0384\\
  $V_{pp\pi}$&   -0.9830  & -0.4231&-0.9875&   -1.1509  & -1.1513 & -1.1146\\
  $\lambda_{a}$ &   0.3943/2  & -0.0159/2&0.5350/2&   0.2083/2  & 0.4423/2& 0.4237/2\\
  $\lambda_{c}$&   0.7216/2  & 0.7651/2& 0.1950/2&   0.2856/2  & 0.1246/2& 0.0306/2\\
   $E_{f}$&   3.32248   & -1.33027 & 3.32331 &   0.173742  & 0.293268 & 1.71368\\
\hline 
\end{tabular}
\caption{Tight-binding parameters (in eV) of different material.}\label{table:1}
\end{table}
\end{center}
\end{widetext}

\begin{figure}
    \centering
    \includegraphics[width=1.0\columnwidth]{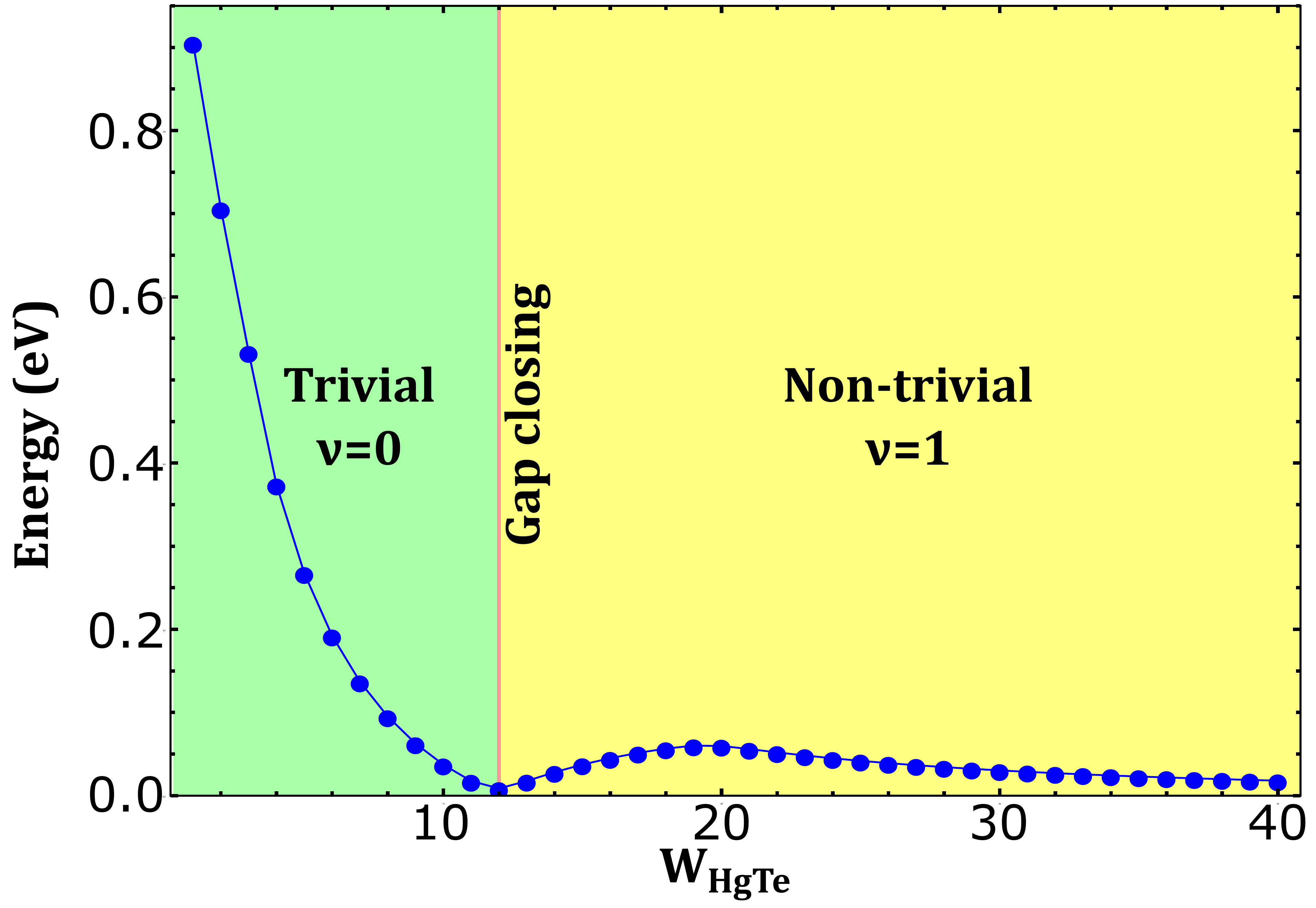}
    \caption{Energy gap $E_{\rm gap}$ and topological invariant $\nu$ as a function of the quantum well thickness $W_{\rm HgTe}$ in CdTe$_{10}$/HgTe$_{W_{\rm HgTe}}$/CdTe$_{10}$ heterostructure. The transition from topologically trivial $\nu=0$ to nontrivial $\nu=1$ phase takes place at $W_{\rm HgTe, c}=12$  unit cells. }
    \label{fig:topo_trans}
\end{figure}

The two-dimensional Hamiltonian ${\cal H}^{2D}\left(k_{1},k_{2}\right)$ of a quantum heterostructure $X_{W_{X}}/Y_{W_{Y}}/X_{W_{X}}$, consisting of $W_X$ unit cells of material $X$ (insulating barrier), $W_Y$ unit cells of material $Y$ (quantum well) and $W_X$ unit cells of material $X$ (insulating barrier), stacked along $\textbf{n}_3$ direction can be written as
\begin{eqnarray}
{\cal H}^{2D}\left(k_{1},k_{2}\right)&=&\sum_{i} |i\rangle\langle i|\otimes H_{0}\left(k_{1},k_{2}, i\right) \nonumber \\&& \hspace{-1.2cm}+ \left( \sum_{i} |i\rangle\langle i+1|\otimes H_{1}\left(k_{1},k_{2}, i\right) + {\rm H.c.}\right),
\end{eqnarray}
where $|i \rangle$ is the basis state for the $i$th unit cell along the $\textbf{n}_3$ direction, and 
\begin{equation}
H_0(k_1, k_2, i)= \begin{cases} H_0^X, & 0< i \leq W_X  \\
H_0^y, & W_X < i \leq W_X+W_Y \\
H_0^X, & W_X+W_Y < i \leq 2 W_X +W_Y
\end{cases},
\end{equation}
\begin{equation}
H_1(k_1, k_2, i)= \begin{cases} H_1^X, & 0< i < W_X  \\
H_1^{XY}, & i= W_X  \\
H_1^y, & W_X < i < W_X+W_Y \\
H_1^{XY}, & i= W_X+W_Y  \\
H_1^X, & W_X+W_Y < i \leq 2 W_X +W_Y
\end{cases},
\end{equation}
with $H^X_{0}\left(k_{1},k_{2}\right)$ and $H^X_{1}\left(k_{1},k_{2}\right)$ obtained from the Fourier decomposition of  $\mathcal{H}(\textbf{k})$ [Eq.~(\ref{eq:Ham3D})] of material $X$  
\begin{equation}
\mathcal{H}(\textbf{k})=H_{0}\left(k_{1},k_{2}\right)+e^{ik_{3}}H_{1}\left(k_{1},k_{2}\right)+e^{-ik_{3}}H_{1}^{\dagger}\left(k_{1},k_{2}\right).
\end{equation}
We assume that the hopping matrices between the different materials $H_1^{XY}$ can be written as
\begin{equation}
H_1^{XY}=(1-x)\frac{H_1^X+H_1^y}{2}, \label{eq:HXY}
\end{equation}
allowing us to turn the coupling between the materials on and off by changing $x$ continuously from $0$ to $1$. This is useful in the following when we study the microscopic origin of the non-topological edge modes. If not otherwise stated we use $x=0$ so that the coupling is turned on. We have benchmarked the tight-binding model by studying the topological phase transition in CdTe/HgTe/CdTe quantum wells. As shown in Fig.~\ref{fig:topo_trans} we obtain a transition from topologically trivial to nontrivial phase at $W_{{\rm HgTe}, c}=12$ unit cells in approximate agreement with the previous studies \cite{RevZhang, BHZ}. 

\begin{figure}
\includegraphics[width=1.0\columnwidth]{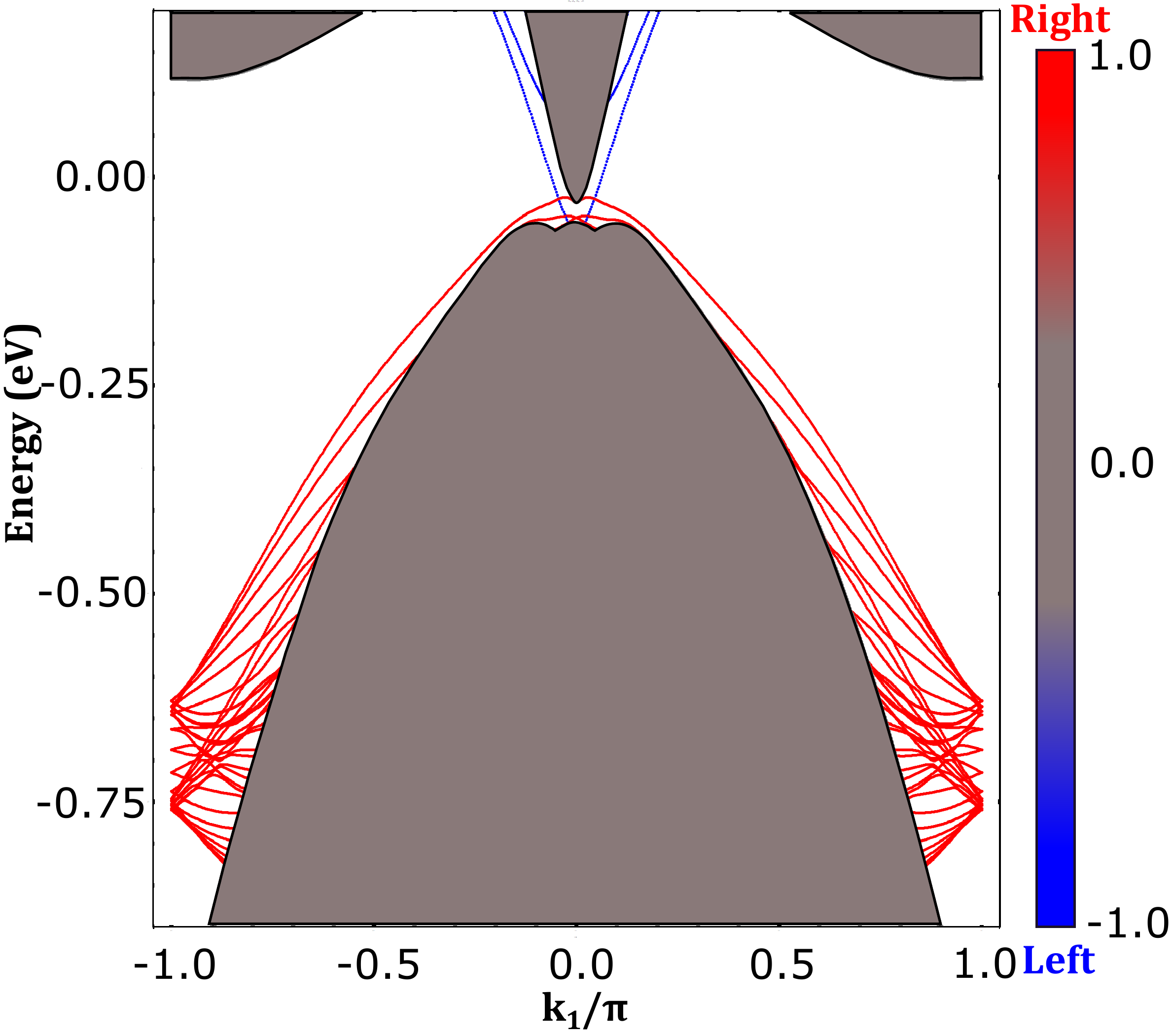}\caption{Edge states spectrum of topologically nontrivial CdTe$_{10}$/HgTe$_{16}$/CdTe$_{10}$ heterostructure with width $W'=300$ unit cells. The colors (normalized to maximum absolute values)  indicate the projection of the eigenstates onto $20$ unit cells  located at the left (blue) and right (red) edge of the system. Two pairs of topological helical edge states connect the conduction and valence band through the bulk gap. Additionally, there exists  large number of non-topological edge states far away from the Fermi level.  \label{fig:2}}
\end{figure}
\begin{figure}
\includegraphics[width=0.93\columnwidth]{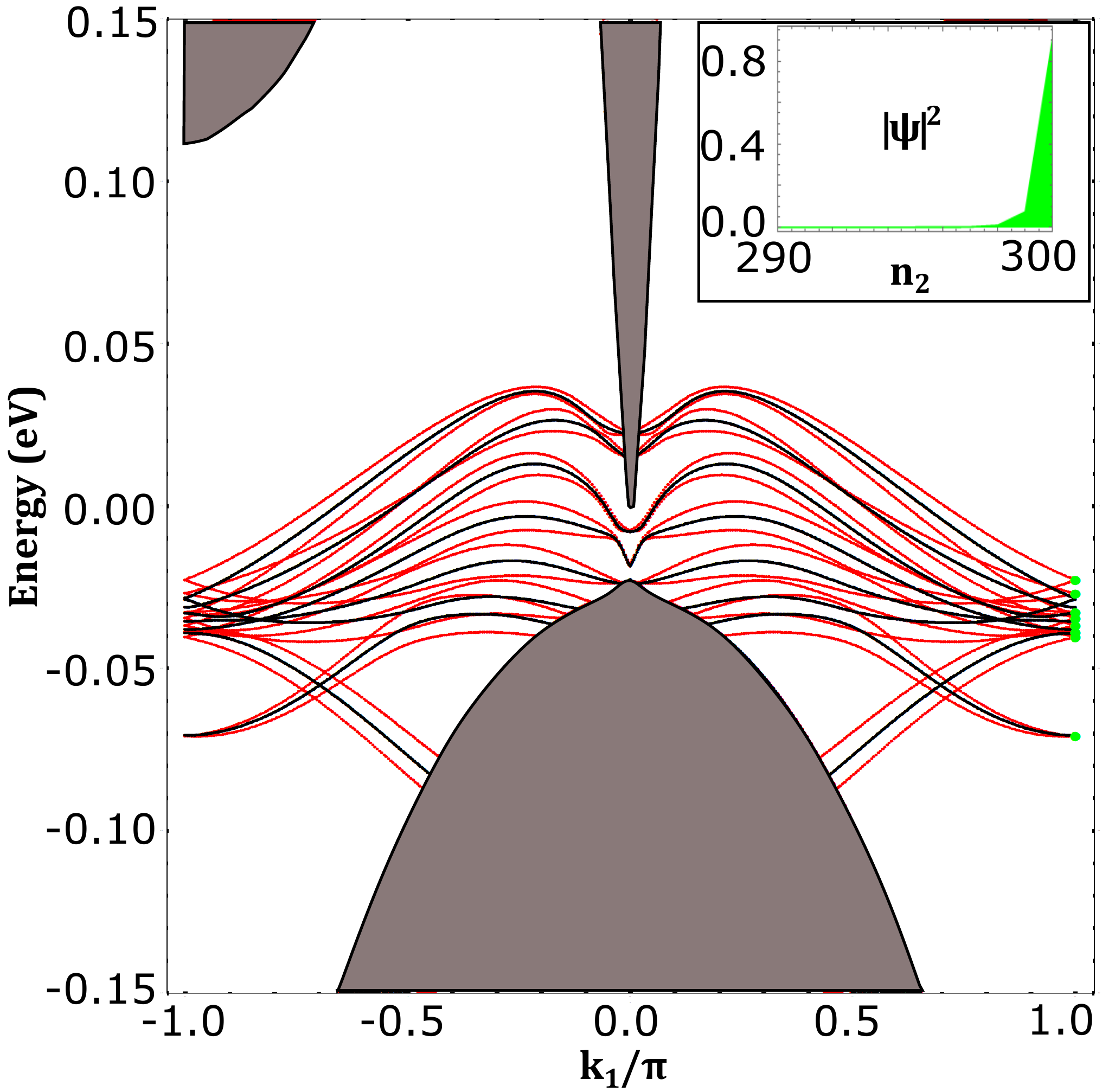}\caption{Edge states spectrum of topologically trivial  CdTe$_{10}$/HgS$_{8}$/CdTe$_{10}$ heterostructure with width $W'=300$ unit cells. In the presence of spin-orbit coupling we have used the same colors as in Fig.~\ref{fig:2} (states localized at the right edge are red), whereas in the absence of spin-orbit coupling ($\lambda_{1,2} = 0$) the projection on the right side is indicated with black. Inset: Local density of states (LDOS) as a function of position $\textbf{n}_2$ close to the right edge for the edge states at $k_1=\pi$ (green dots).\label{fig:3}}
\end{figure}
\begin{figure}
\includegraphics[width=1.0\columnwidth]{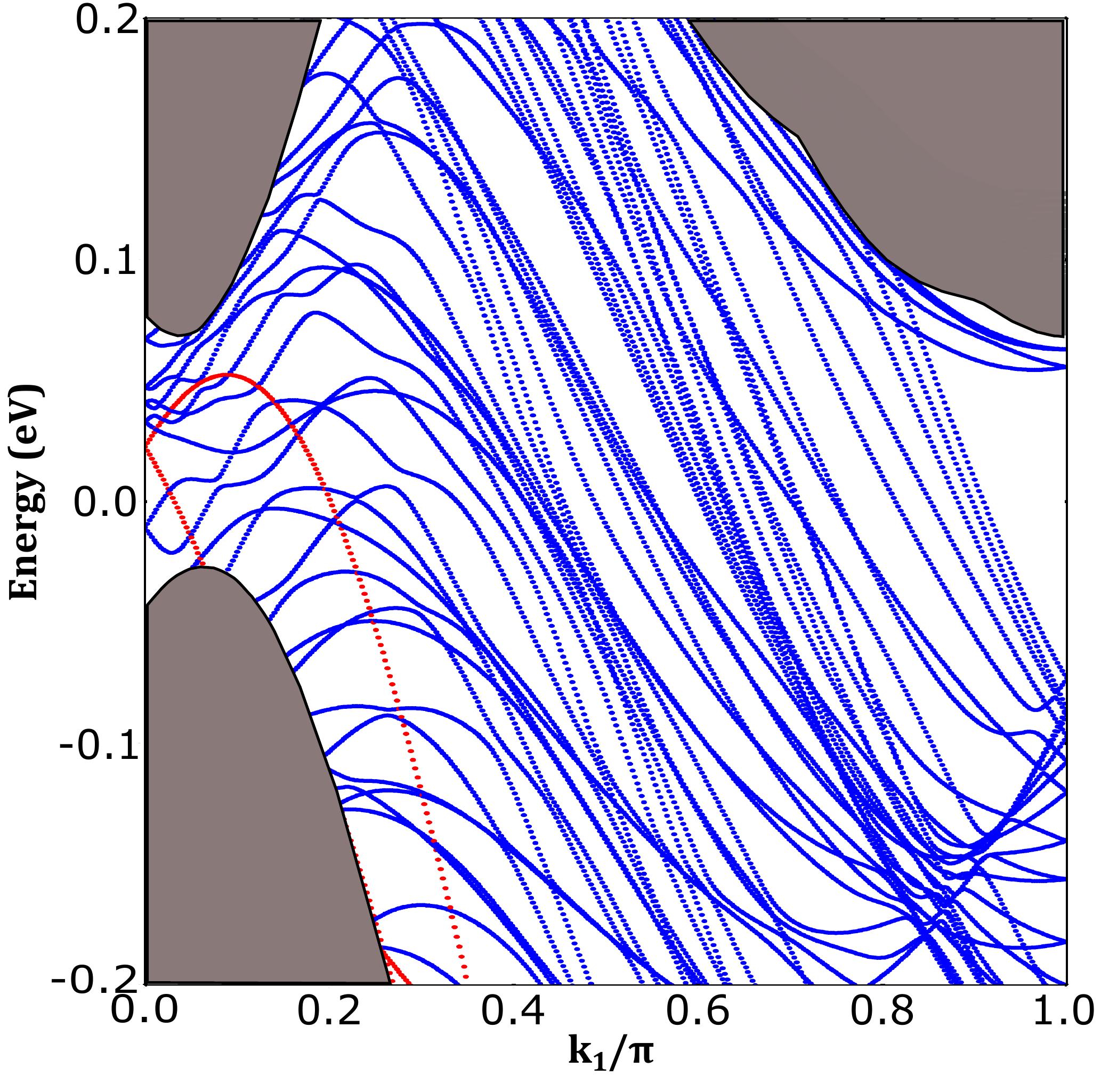}\caption{Edge states spectrum of topologically trivial AlSb$_{10}$/InAs$_{10}$/GaSb$_{10}$/AlSb$_{10}$ heterostructure with width $W'=300$ unit cells. \label{fig:4}}
\end{figure}

To study the edge states spectra of these materials we consider one-dimensional ribbons of width $W'$ along $\textbf{n}_2$ direction 
\begin{equation}
\mathcal{H}^{1D}_{k_{1}}\!=\!\mathbbm{1}_{W^{\prime}}\otimes H^{\prime}_{0}\left(k_{1}\right)\!+\!D\otimes H^{\prime}_{1}\left(k_{1}\right)\!+\!D^{\intercal}\otimes H^{\prime\dagger}_{1}\left(k_{1}\right), \label{ribbon}
\end{equation}
where 
\begin{equation} 
D=\sum_{i=1}^{W^{\prime}-1} |i\rangle\langle i+1|, \label{eq:d}
\end{equation}  
%is $W^{\prime} \times W^{\prime}$ matrix
%\begin{equation}
%D=\begin{pmatrix}0 & 1 & 0 & 0 & 0\\
%0 & 0 & 1 & 0 & 0\\
% &  &  & \ddots\\
%0 & 0 & 0 & 0 & 1\\
%0 & 0 & 0 & 0 & 0
%\end{pmatrix},\label{eq:d}
%\end{equation}
and $H^{\prime}_{0}\left(k_{1}\right)$ and $H^{\prime}_{1}\left(k_{1}\right)$ are obtained from the Fourier decomposition 
\begin{equation}
\mathcal{H}^{2D}_{k_{1},k_{2}}=H^{\prime}_{0}\left(k_{1}\right)+e^{ik_{2}}H^{\prime}_{1}\left(k_{1}\right)+e^{-ik_{2}}H^{\prime\dagger}_{1}\left(k_{1}\right).
\end{equation}
Based on the previous studies  we expect that CdTe/HgTe/CdTe quantum wells support a pair of counterpropagating helical edge states connecting through the bulk gap in the topologically nontrivial regime $W_{\rm HgTe}>W_{{\rm HgTe}, c}$ \cite{RevZhang,BHZ,QSH_HgTe} whereas we expect that there are no edge states in the trivial regime $W_{\rm HgTe}<W_{{\rm HgTe}, c}$.  However, we find that the spectra of CdTe/HgTe/CdTe and CdTe/HgS/CdTe  compounds support additional edge states as shown in Figs.~\ref{fig:2} and \ref{fig:3}, respectively.  In the case of CdTe/HgTe/CdTe the additional edge states appear at energies far away from the bulk gap so that they do not contribute to the transport but in CdTe/HgS/CdTe the edge states are observed inside the bulk gap giving rise to the possibility of multimode edge transport. We find that these additional edge states can appear in nontrivial and trivial heterostructures, and in  Fig.~\ref{fig:3} we also demonstrate that the  spin-orbit coupling does not play a significant role in the appearance of the non-topological edge states. Finally, in Fig.~\ref{fig:4} we demonstrate that this type of non-topological edge states appear also in the topologically trivial AlSb/InAs/GaSb/AlSb heterostructures inside the bulk gap giving rise to the possibility of multimode edge transport. In our calculations the non-topological edge modes are mostly localized on one of the edges because we have used particular edge terminations, determined by Eq.~(\ref{ribbon}), in construction of the ribbons. The dependence on the termination highlights the non-topological nature of the edge modes, but in real materials the edges are not expected to be perfectly ordered, and therefore we expect that the  non-topological edge modes are distributed on both edges.  The rest of the paper is devoted to the understanding of the microscopic origin and the other properties of the  non-topological edge states. 

\section{Minimal model \label{sec:minimal}}

In order to understand the microscopic origin of the non-topological edge modes, in this section we consider a model for a buckled honeycomb lattice of anions and cations, which can be considered to be the minimal building block for constructing the HgTe/CdTe, HgS/CdTe and InAs/GaSb heterostructures. We note that this minimal model supports flat bands with nontrivial quantum geometry that gives rise to polarization charges at the edges \cite{Rhi17}. In Sec.~\ref{Sec_evominimal} we demonstrate that the polarization charges transform into the additional edge states  when the flat bands are coupled to each other and to the other states to form the Hamiltonian describing the full heterostructure.

\begin{figure}[!h]
\includegraphics[width=1.0\columnwidth]{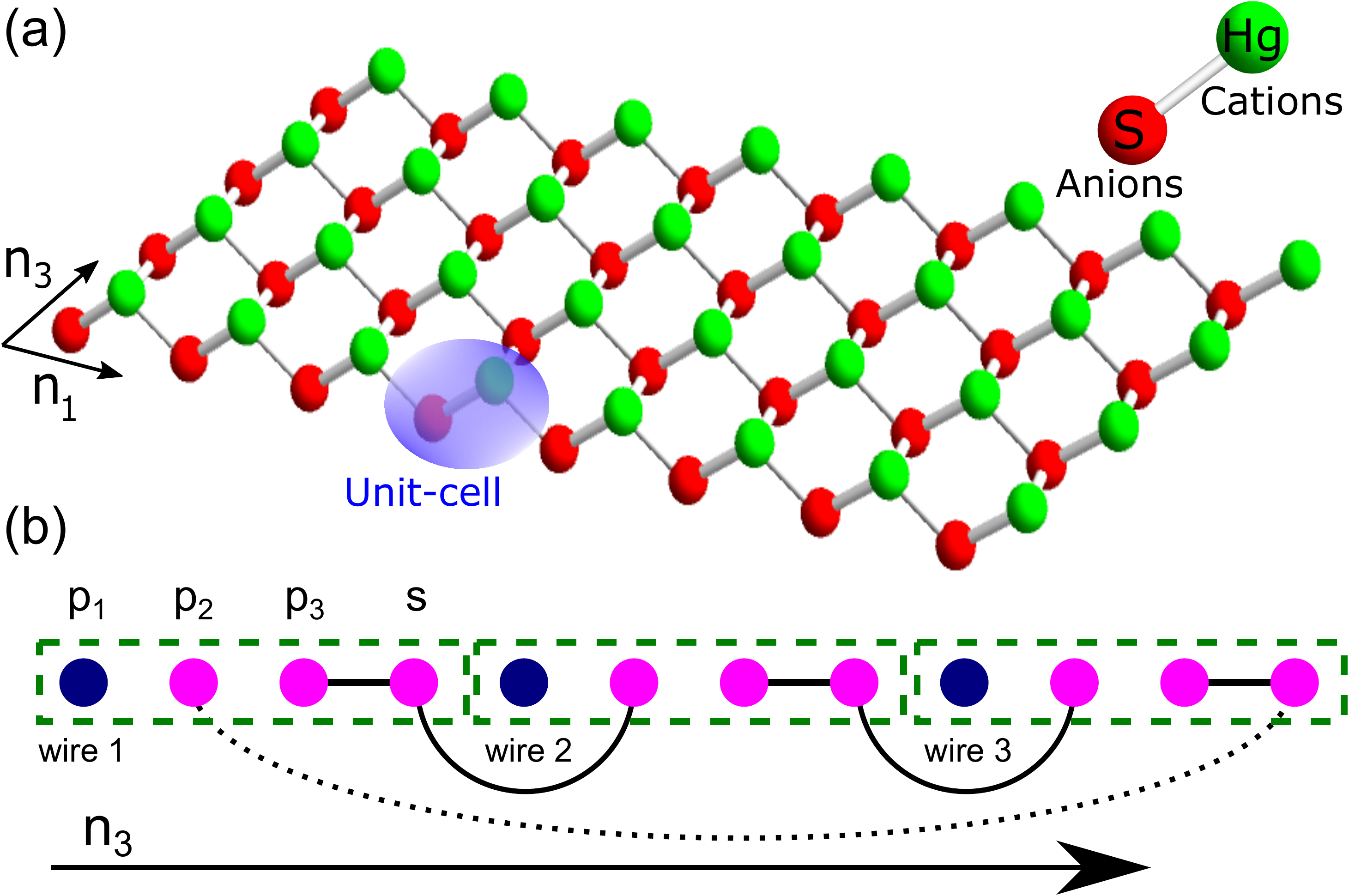}\caption{
(a) Buckled honeycomb lattice of  anions (red) and cations (green) describing the minimal model. The directions in this two-dimensional system correspond to the $n_1$  and $n_3$ in the original three-dimensional model [Eq.~(\ref{eq:Ham3D})]. 
(b) The Zak phases of the flat bands of the model give rise to an  accumulated end charge proportional to  the width of the ribbon $W$  because of an additional symmetry that allows to decompose the Hamiltonian into diagonal blocks. Each dashed rectangle represents
four states, labeled as $p_{1}$, $p_{2}$,
$p_{3}$ and $s$ due to their orbital contents, of a single zigzag chain. The states represented
by purple (blue) disks form $3\times3$ ($1\times1$) diagonal blocks after the decomposition. The black
lines indicate that the states connected by them go to the same block.
The dashed black line denotes a coupling, which is present only in the cylinder geometry with periodic boundary conditions in the transverse
direction.
\label{fig:1b}}
\end{figure}

The minimal model can obtained from the full three-dimensional Hamiltonian [Eq.~(\ref{eq:Ham3D})] by projecting the model to $s$-orbitals of cations and $p$-orbitals of anions, neglecting the spin-orbit coupling terms and setting $k_2=\pi$. Furthermore we set $E_f = 0$ for simplicity. This way we obtain a buckled honeycomb lattice shown in Fig.~\ref{fig:1b}(a). The  2D bulk Hamiltonian of this system is 
\begin{eqnarray}
{\cal H}^{2D}\left(k_{1},k_{3}\right) & = & H_{\parallel}\left(k_{1}\right)+\left(e^{ik_{3}}H_{\perp}+H.c.\right),\label{eq:pbc}
\end{eqnarray}
and 1D Hamiltonian of $W$ unit cells wide ribbon is 
\begin{eqnarray}
{\cal H}^{1D}\left(k_{1}\right) & = & \mathbbm{1}_{W}\otimes H_{\parallel}\left(k_{1}\right)+\left(D\otimes H_{\perp}+H.c.\right),\label{eq:obc}
\end{eqnarray}
where
\begin{eqnarray*}
\frac{H_{\parallel} \left(k_{1}\right)}{E_0} & = & \begin{pmatrix}\eta_{s} & e^{ik_{1}}\!-\!1 & -e^{ik_{1}}\!+\!1 & -e^{ik_{1}}\!-\!1\\
e^{-ik_{1}}\!-\!1 & \eta_{p} & 0 & 0\\
-e^{-ik_{1}}\!+\!1 & 0 & \eta_{p} & 0\\
-e^{-ik_{1}}\!-\!1 & 0 & 0 & \eta_{p}
\end{pmatrix}, 
\end{eqnarray*}
\begin{eqnarray*}
\frac{H_{\perp}}{E_0} & = & \begin{pmatrix}0 & 2 & 2 & 0\\
0 & 0 & 0 & 0\\
0 & 0 & 0 & 0\\
0 & 0 & 0 & 0
\end{pmatrix}, E_0=\frac{V_{s_c p_a\sigma}}{\sqrt{3}}, \eta_{s}=\frac{V_{s_c}}{E_0}, \eta_{p}=\frac{V_{p_a}}{E_0}. 
\end{eqnarray*}
Here $D$ is 
a $W\times W$ matrix of the form (\ref{eq:d}).
If we assume periodic boundary condition in the transverse direction, corresponding to a cylinder geometry instead of a ribbon, the matrix $D$ is replaced by 
%\begin{equation}
%T=\begin{pmatrix}0 & 1 & 0 & 0 & 0\\
%0 & 0 & 1 & 0 & 0\\
% &  &  & \ddots\\
%0 & 0 & 0 & 0 & 1\\
%1 & 0 & 0 & 0 & 0
%\end{pmatrix}.
%\end{equation}
\begin{equation} 
T=\sum_{i=1}^{W-1} |i\rangle\langle i+1|+|W\rangle\langle 1|
= D + |W\rangle\langle 1|. \label{eq:Tpbc}
\end{equation}

The spectrum of ${\cal H}\left(k_{1},k_{3}\right)$ is flat in the direction of $k_{3}$ and there
are two completely flat bands 
\begin{eqnarray}
E_{1,2}\left(k_{1}\right) & = & \frac{1}{2}\left(\eta_{p}\!+\!\eta_{s}\! \mp \!\sqrt{56\!+\!\left(\eta_{p}\!-\!\eta_{s}\right)^{2}\!-\!8\cos k_{1}}\right),\nonumber \\
E_{3}&=&E_{4}=\eta_{p}.
\label{eq:E12}
\end{eqnarray}
According to our knowledge the Hamiltonian ${\cal H}\left(k_{1},k_{3}\right)$ is topologically trivial in all classifications.
Nevertheless, we find  end states, evolving from the dispersive states $\left|E_{1}\left(k_{1}\right)\right\rangle $
and $\left|E_{2}\left(k_{1}\right)\right\rangle $, when the system is finite along the $k_1$ direction.
The end states are localized on one end of the system and 
they can be constructed analytically using the non-Bloch waves ansatz described in
Ref.~\cite{Yok19} (see Appendix \ref{app:end-states}).
The charge density of the end-states has a decay length 
\begin{equation}
\xi=\frac{1}{2\ln3}
\end{equation}
and the energies of the end-states are given by 
\begin{equation}
E_{1,2}\left(-i \ln 3\right) = \frac{1}{2}\left(\eta_{p}\!+\!\eta_{s}\! \mp \!\sqrt{\frac{128}{3}\!+\!\left(\eta_{p}\!-\!\eta_{s}\right)^{2}}\right). \label{end-state-energy}
\end{equation}
Importantly, we obtain these end states with the same energies $E_{1(2)}$  and localization length $\xi$ for all values of $k_3$ so that the number of end modes is proportional to the width $W$ of the ribbon. The existence of the end modes depends on the lattice termination, so that with the termination used in Appendix \ref{app:end-states} all of the end modes are localized at the right end of the system.

The non-Bloch waves ansatz \cite{Yok19} tells us that there cannot be  end states
evolving from the flat bands $E_3$ and $E_4$, because the energies generically must be of the form $E_{n}(q)$ with some complex
$q$, so that the flat bands cannot
give rise to a state of energy different than $\eta_{p}$. However, the  flat bands can still lead to a charge accumulation
at the ends of the system due to the Zak phase \cite{Rhi17}. Typically, such kind of quantum geometric effect on the charge accumulation is small because the Zak phase is only defined modulo $2\pi$ but we find that in our system the accumulated charge scales with the width $W$ of the ribbon because of an  additional symmetry of the system. 

\begin{figure}
\includegraphics[width=1.0\columnwidth]{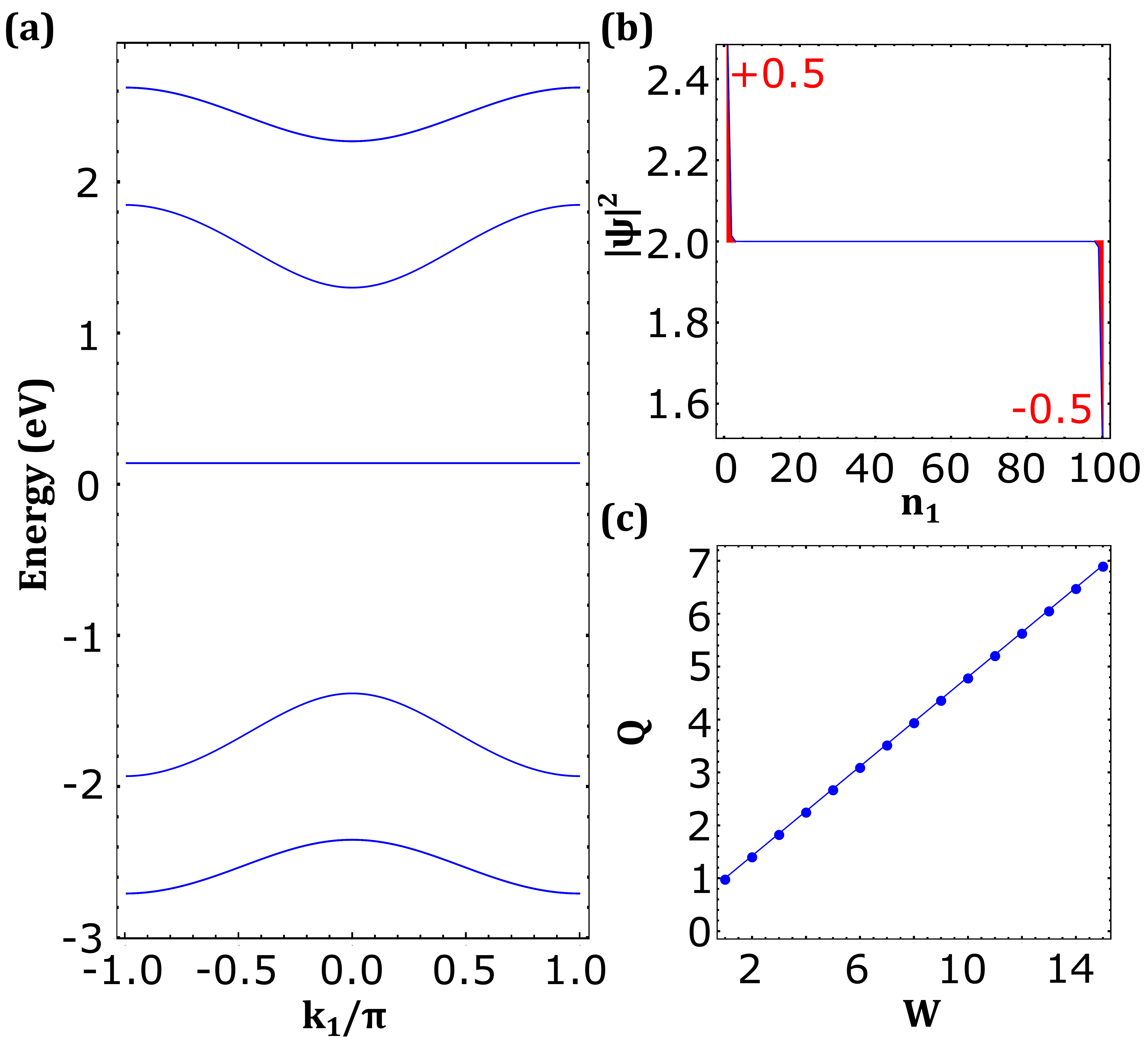}\caption{(a) Band structure of minimal model ribbon of width  $W=8$ unit cells in the $\textbf{n}_3$  direction. The flat bands have a degeneracy $16$, whereas the highest and lowest dispersive bands $E_{1,2}(k_1)$ are 7-fold degenerate and the other dispersive bands $E_{\parallel 1,2}(k_1)$ are non-degenerate.
(b) LDOS of flat bands  for a single zigzag chain of length $L=100$ unit cells as a function of position  $\textbf{n}_1$ with the charge accumulate quantized at both end. (c) Accumulated charge as  a function of thickness $W$ along the $\textbf{n}_3$  direction. We have used $V_{sc}=-1.5535, V_{pa}=-1.1909, V_{s_c p_a}=1.1517, E_f=-1.33027$ (in meV) corresponding to HgS.    \label{fig:8}}
\end{figure}

Namely, we can transform the ribbon Hamiltonian (\ref{eq:obc}) [and the cylinder Hamiltonian with periodic boundary conditions (\ref{eq:Tpbc})] as 
\begin{equation}
{\cal H}^{ 1D}(k_{1})\to{\cal U}^{\dagger}{\cal H}^{1D}(k_{1}) \, {\cal U}, \quad {\cal U}=\mathbbm{1}_{W}\otimes U,  
\end{equation}
where 
\begin{eqnarray}
U &=& \begin{pmatrix}0 & 0 & 0 & 1\\
 \frac{\cos\frac{k_{1}}{2}}{\sqrt{3-\cos k_{1}}} & \frac{1}{\sqrt{2}} &  \frac{-i\sin\frac{k_{1}}{2}e^{-i\frac{k_{1}}{2}}}{\sqrt{\frac{3-\cos k_{1}}{2}}} & 0\\
 \frac{-\cos\frac{k_{1}}{2}}{\sqrt{3-\cos k_{1}}} & \frac{1}{\sqrt{2}} &  \frac{i\sin\frac{k_{1}}{2}e^{-i\frac{k_{1}}{2}}}{\sqrt{\frac{3-\cos k_{1}}{2}}} & 0\\
 \frac{2i\sin\frac{k_{1}}{2}}{\sqrt{3-\cos k_{1}}} & 0 &  \frac{-\cos\frac{k_{1}}{2} e^{-i\frac{k_{1}}{2}}}{\sqrt{\frac{3-\cos k_{1}}{2}}} & 0 \end{pmatrix}, 
\end{eqnarray}
leading to $W$ identical $1\times1$ blocks 
\begin{equation}
B_1=E_0 (\eta_p),
\end{equation}
$W-1$ identical $3 \times 3$ blocks  
\begin{eqnarray}
B_{3} & = & E_0 \begin{pmatrix}\eta_{p} & \sqrt{6-2\cos k_1} & 0\\
\sqrt{6-2\cos k_1} & \eta_{s} & \sqrt{8}\\
0 & \sqrt{8} & \eta_{p}
\end{pmatrix},
\end{eqnarray}
and one $3 \times 3$ block 
\begin{eqnarray}
B'_{3} & = & E_0 \begin{pmatrix}\eta_{p} & \sqrt{6-2\cos k_1} & 0\\
\sqrt{6-2\cos k_1} & \eta_{s} & c\sqrt{8}\\
0 & c\sqrt{8} & \eta_{p}
\end{pmatrix},\nonumber 
\end{eqnarray}
where $c=0$ ($c=1$) for the ribbon (cylinder) geometry. A schematic view of how the different states, represented by the columns of matrix $U$,  contribute to each block is shown in Fig.~\ref{fig:1}(b). Each of the
$B_{3}$ blocks contains one flat band with energy $\eta_{p}$ and
two bands with energies $E_{1,2}\left(k_{1}\right)$ given in Eq.~(\ref{eq:E12}). In the cylinder geometry with $c=1$ block $B'_{3}$ is identical to $B_{3}$
but in the ribbon geometry with $c=0$ the block $B'_{3}$ contributes a flat band with energy $\eta_{p}$
and two dispersive bands with energies
\begin{equation}
E_{\parallel 1,2}\left(k_{1}\right)=\frac{1}{2}\left(\eta_{p}\!+\!\eta_{s}\! \mp \!\sqrt{24\!+\!\left(\eta_{p}\!-\!\eta_{s}\right)^{2}\!-\!8\cos k_{1}}\right).
\label{eq:E12p}
\end{equation}
The energy bands of a ribbon with $c=0$ are shown in Fig.~\ref{fig:8}(a).
Every $B_{3}$ block and $B_3'$ block with $c=1$ contributes 
two end states with energies $E_{1,2}(-i\ln3)$ if the system is opened in  the $k_1$ direction, whereas $B'_{3}$ with $c=0$ contributes two end states with energies $E_{\parallel 1,2}(-i\ln3)$.

Because of the block decomposition the accumulated charge at the end of the ribbon, when it is opened in the $k_1$ direction, is related to the sum the Zak phases of the different blocks.  
From an explicit calculation, using the standard prescription
\begin{equation}
\gamma_{p}=\frac{1}{i}\int_{0}^{2\pi}\left\langle E_{p}\left(k\right)\right|\partial_{k}\left|E_{p}\left(k\right)\right\rangle 
\end{equation}
and the analytical form of the eigenvectors, 
we get that the Zak phases of the flat bands from the blocks $\left\{ B_{1},B_{3},B'_{3}\right\}$ are
\begin{equation}
\gamma_{1} =  \pi, \ 
\gamma_{3}  =  -\frac{\pi}{\sqrt{3}},\ 
\gamma'_{3} =  -c\frac{\pi}{\sqrt{3}}.
\label{Zak-phases}
\end{equation}
Thus, inspired by Ref.~\cite{Rhi17},
we expect that the difference of the left and right boundary
charges  due to the geometric phases of the flat bands is 
\begin{equation}
Q=\frac{1}{\pi}\left(W\gamma_{1}+\left[W-1\right]\gamma_{3}+\gamma'_{3}\right). \label{Q-Zak}
\end{equation}
Indeed, we numerically find that 
\begin{equation}
Q =  \begin{cases} \frac{1}{\sqrt{3}}+W\left(1-\frac{1}{\sqrt{3}}\right), \ c=0 \\
W\left(1-\frac{1}{\sqrt{3}}\right), \ c=1 \end{cases},
\end{equation}
demonstrating that Eq.~(\ref{Q-Zak}) correctly describes the accumulated boundary charge, which increases proportionally to the width $W$ of the system [see Fig.~\ref{fig:8}(b),(c)]. 

Note that the above considerations are valid also when we set $k_1=\pi$ as a starting point and consider a 2D system in the $k_2$---$k_3$ plane leading to a 1D ribbon or cylinder Hamiltonian ${\cal \tilde{H}}^{1D}\left(k_{2}\right)$ analogical to Eq.~(\ref{eq:obc}). Despite of both Hamiltonians being apparently quite different we show in Appendix \ref{app:duality} that they are related by a unitary transformation in case of the cylinder geometry and differ only in block $B_3^{\prime}$ in case of the ribbon geometry. In the former case the end states and the number of flat bands are the same as for ${\cal H}^{1D}\left(k_{1}\right)$ whereas in the latter one the number of edge states (flat bands) is smaller by two (larger by two) because of the difference between $B_3^{\prime}$ blocks, see Appendix \ref{app:duality}.

\section{Connection between the non-topological edge states and the flat bands in the  minimal model} \label{Sec_evominimal}

Next we demonstrate that the non-topological edge states indeed originate from the flat bands of the minimal model by 
interpolating between the Hamiltonians and following the evolution of the edge states. 

\begin{figure}
\includegraphics[width=1.0\columnwidth]{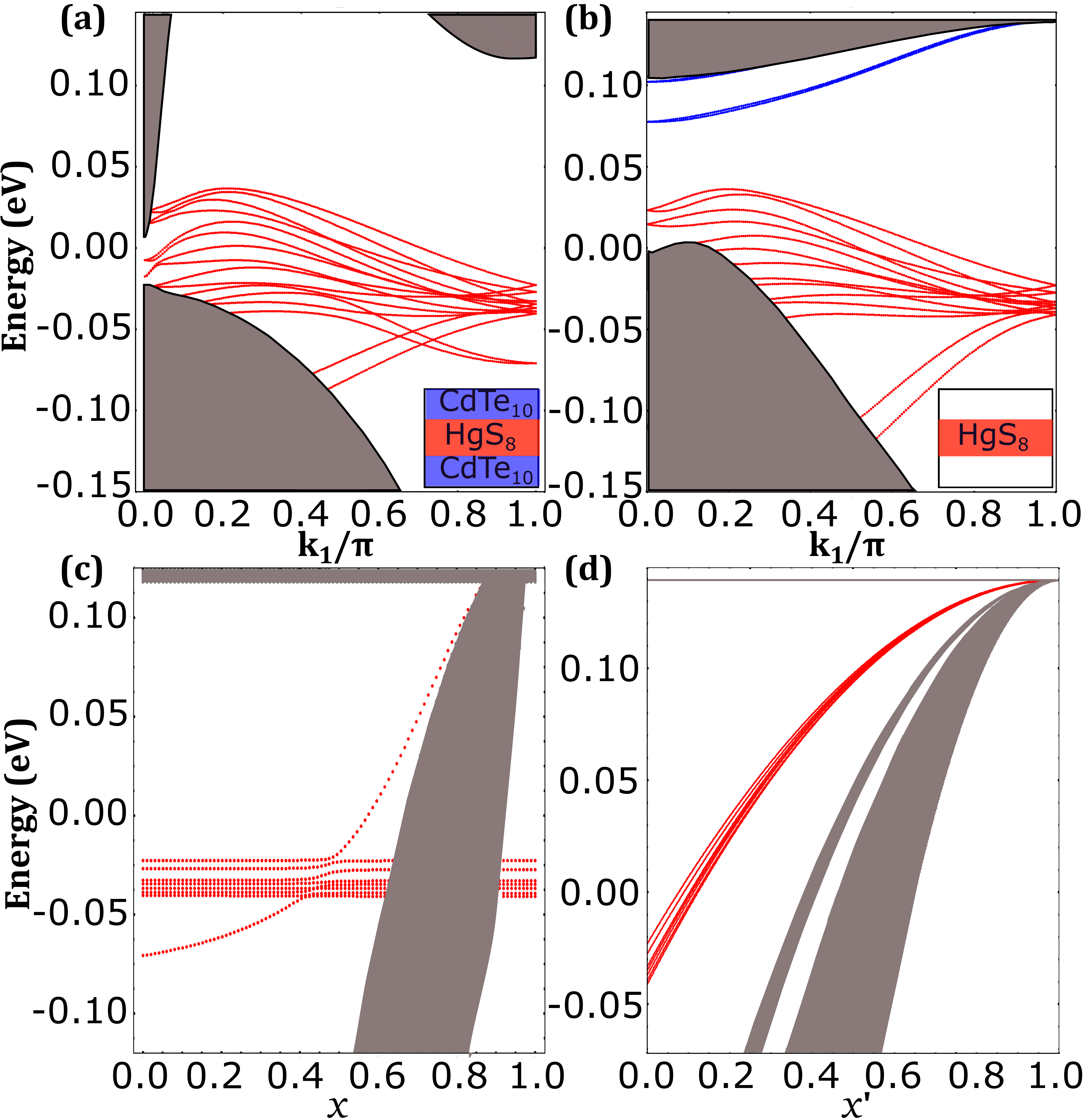}\caption{(a),(b) Edge states spectra of CdTe$_{10}$/HgS$_{8}$/ CdTe$_{10}$ and HgS$_{8}$ of width $W'=300$ unit cells in $\textbf{n}_2$  direction, respectively. 
(c) Evolution of eigenenergies at  $k_{1}=\pi$ as function of $x$ interpolating between coupled ($x=0$) and uncoupled ($x=1$) systems of HgS$_8$ and CdTe$_{10}$ as described in Eq.~(\ref{eq:HXY}). (d) Evolution of eigenenergies at $k_{1}=\pi$ from  $HgS_{8}$ ($x'=0$) to the minimal model ($x'=1$).   \label{fig:5}}
\end{figure}

We first notice that similar non-topological edge states are obtained both in the full heterostructure of CdTe$_{10}$/HgS$_{8}$/ CdTe$_{10}$ and in  HgS$_{8}$  [Fig.~\ref{fig:5}(a),(b)].  Indeed, by interpolating between coupled $x=0$ and uncoupled $x=1$ systems of HgS$_8$ and CdTe$_{10}$, as described in Eq.~(\ref{eq:HXY}), we find that the majority of the edge states
at high symmetry point $k_{1}=\pi$ remains unchanged 
throughout the
evolution [Fig.~\ref{fig:5}(c)], despite the fact that the coupling has a large impact on the bulk state energies. Moreover, by studying the eigenvectors we conclude that the edge states are located in the HgS system. 
Therefore, we conclude that the CdTe barriers are not important for understanding the additional edge modes.

We can further trace  back the origin of additional edge states from the HgS$_8$ system to the minimal model by interpolating the model parameters  
\begin{equation}
\vec{P}(x^\prime) =\left( 1-x^{\prime} \right) \vec{P}_{\rm HgS}+ x^{\prime} \vec{P}_{\rm min}
\label{eq:inter}     
\end{equation}
between the HgS parameters 
\begin{eqnarray}
\vec{P}_{\rm HgS}&=&\bigg\{V_{s_a},V_{s_c}V_{p_a},V_{p_c},V_{ss\sigma},  V_{s_a p_c\sigma},V_{s_c p_a\sigma}, \nonumber \\
 && \hspace{1cm} V_{pp\sigma}, V_{pp\pi}, \lambda_{a},\lambda_{c},E_{f} \bigg\}
\end{eqnarray}
and the minimal model parameters
\begin{equation}
\vec{P}_{\rm min}=\left\{ V_{s_a},V_{s_c},V_{p_a},V_{p_c},0,V_{s_c p_a\sigma}, V_{s_a p_c\sigma}, 0,0, 0,0,E_{f}\right\}.
\end{equation}
By following the evolution of the edge states of HgS ($x^\prime=0$) to the minimal model ($x^\prime=1$) we find that the edge states are indeed connected to the flat bands of the minimal model [see Fig.~\ref{fig:5}(d)].

\begin{figure}[!h]
\includegraphics[width=1.0\columnwidth]{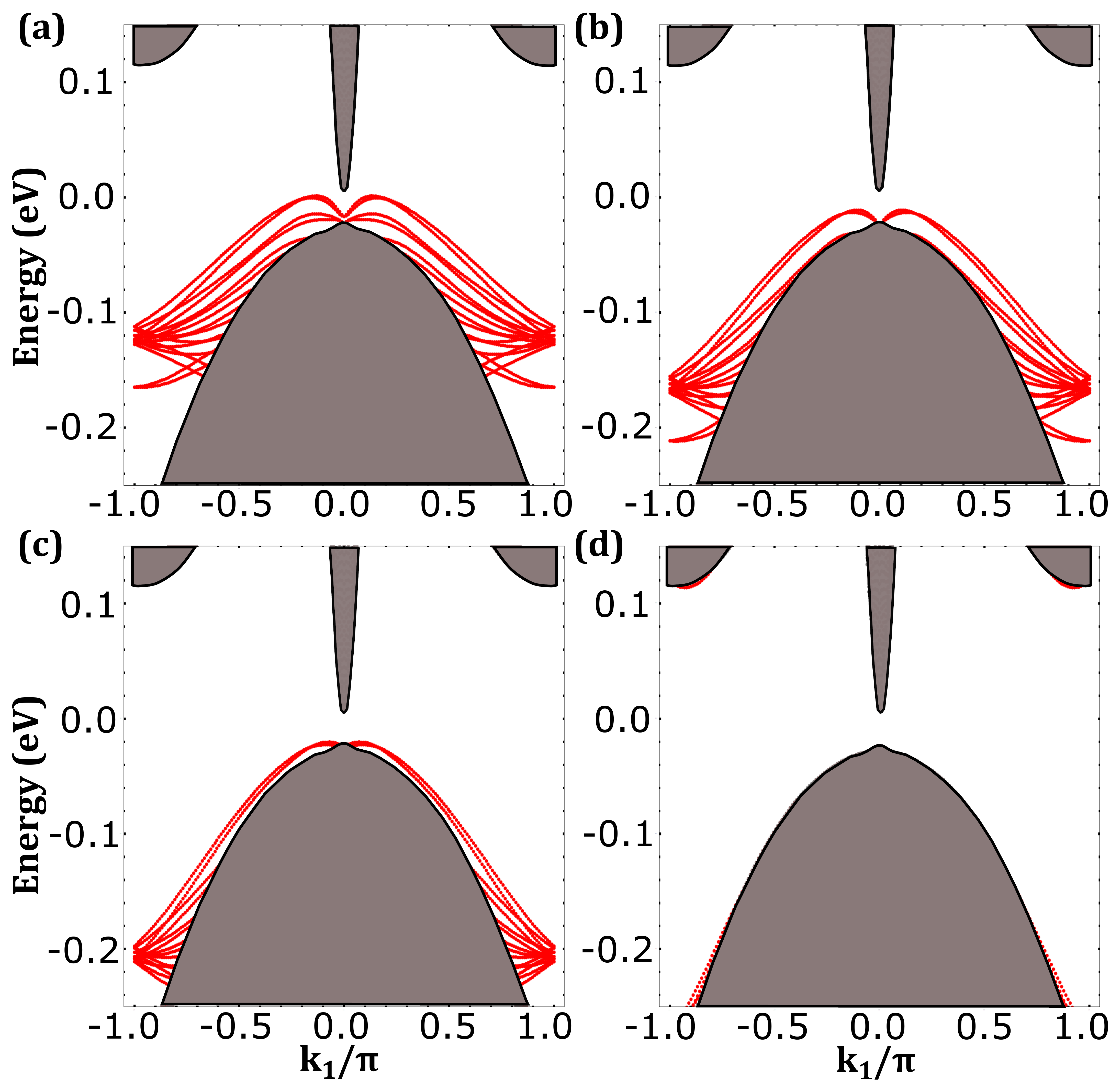}\caption{Edge states  spectra of CdTe$_{10}$/HgS$_{8}$/CdTe$_{10}$  of width $W'=300$ unit cells with an on-site potential $\delta$ applied on the lattice sites at the right edge of the system. The values of on-site potential are (a) $\delta = -0.1 eV$, (b) $\delta = -0.15 eV$, (c) $\delta = -0.2 eV$, and (d) $\delta = -0.3 eV$. \label{fig:6}}
\end{figure}

\section{Effect of edge potential on the non-topological edge states}

We have argued that the additional edge states are not topological. This suggests that it should be possible to remove them from the energy gap between the valence and conduction bands by applying an edge potential for example with the help of a side gate.  In Fig.~\ref{fig:6}
we show the edge state spectra of   CdTe$_{10}$/HgS$_{8}$/CdTe$_{10}$ in the presence of additional  on-site potential $\delta$ applied on the lattice sites at the right edge of the system. By decreasing the value of $\delta$ we find that all non-topological edge states can indeed be removed from the bulk gap. 

\section{Conclusions}

We have shown that HgTe/CdTe, HgS/CdTe and InAs/GaSb heterostructures  support additional edge states which are sensitive to the edge termination, and we have  traced the microscopic origin of these states back to a minimal model supporting flat bands with a nontrivial quantum geometry that gives rise to polarization charges at the edges. Non-topological edge states have been observed in quantum spin Hall insulator candidate materials deteriorating the quality of the quantum spin Hall effect.  Importantly, our results suggest that these states can be removed from the bulk energy gap  by modifying the edge potential for example with a side gate or chemical doping.  

\begin{acknowledgments}

The research was partially supported by the Foundation for Polish Science through the IRA Programme co-financed by EU within SG OP. T.H.  acknowledges
the computational resources provided by
the Aalto Science-IT project and the 
financial support from the 
Academy of Finland Project No.
331094.
W.B. acknowledges support by Narodowe Centrum Nauki (NCN, National Science Centre, Poland) Project No. 2019/34/E/ST3/00404.
R. I. and C. A. acknowledge support by the National Science Center in the framework of the "PRELUDIUM" (Decision No.: DEC-2020/37/N/ST3/02338).
We acknowledge the access to the computing facilities of the Interdisciplinary
Center of Modeling at the University of Warsaw, Grants  G75-10, G84-0 and GB84-1.
We acknowledge the CINECA award under the ISCRA initiatives  IsC93 "RATIO" and IsC99 "SILENTS" grant, for the availability of high-performance computing resources and support.

\end{acknowledgments}

\appendix

\section{Computational details of the density functional calculations \label{app:DFT}}

We performed electronic structure 
calculations by using the VASP \cite{VASP2} package based
on plane wave basis set and projector augmented wave method \cite{VASP}.
A plane-wave  energy cutoff of 250~eV has been used. We have performed the calculations using 8$\times$8$\times$8 k-point Monkhorst-Pack grid in presence of SOC with 512 k-points and in the absence of SOC with 176 k-points in the independent Brillouin zone. For the treatment of exchange-correlation, the modified Becke-Johnson exchange potential together with local density approximation for the correlation potential scheme \cite{MBJLDA1,MBJLDA2} has been considered.   
In particular, we have an improvement in the band gap \cite{Camargo12}, and consequently an improvement of the spin-orbit splitting close to the gap. Similar settings for the DFT calculations were used to describe HgTe and InAs \cite{Islam22,Hussain22}.

After obtaining the Bloch wave functions $\psi_{n,\textbf{k}}$, the Wannier functions \cite{Marzari97,Souza01} are built up using the Wannier90 code \cite{Mostofi08}.
To extract the low energy properties of the electronic bands, we have used the Slater-Koster interpolation scheme to obtain a long-range tight-binding model with the Wannier function method \cite{Mostofi08}.
To obtain the first-neighbour tight-binding model, we require
the tight-binding model to fit the DFT band structure at high-symmetry points $\Gamma$, X and L.
Regarding the spin-orbit, we extracted the effective SOC of the anion $\lambda_a$ from the formula  
$3\lambda_a$=E($\Gamma_8$)-E($\Gamma_7$) where $\Gamma_8$ and $\Gamma_7$ anion energy levels were obtained from first-principle calculations \cite{Autieri21}, the same was done for the cation $\lambda_c$.
Note that in Refs. \cite{Autieri21} we used different notation where $\frac{3\lambda_a}{2}$=E($\Gamma_8$)-E($\Gamma_7$).
Following this procedure, we obtained the hopping parameters, the on-site energies and the spin-orbit constants. 
The experimental lattice constants, which coincide with those used in our DFT calculation except for HgS, are reported in Table \ref{lattice_constants}.
For HgS we have used a larger lattice constant $a_{0}$=6.835 {\AA} to obtain that the effective SOC is close to zero.

\begin{table}[h!] 
	\begin{centering}
		\begin{tabular}{|c|c|c|c|c|c|c|}
			\hline
          & HgTe\cite{madelung1982numerical}  & HgS\cite{madelung1982numerical} & CdTe\cite{madelung1982numerical} & InAs\cite{Vurgaftman01} & GaSb\cite{Vurgaftman01} & AlSb\cite{Vurgaftman01}   \\ \hline
$a_{0}$ & 6.460 & 5.851 & 6.480 & 6.050 & 6.082 & 6.128 \\ \hline
	    \end{tabular}
	\end{centering}
	\caption{Experimental lattice constant at 8K for HgTe and at 300K for HgS, CdTe, at 0 K extrapolation for InAs, GaSb and AlSb. All the values are in {\AA}.}
	\label{lattice_constants}
\end{table} 

We demonstrate why the effective spin-orbit $\lambda_a$ is negative for the p-state of S in HgS \cite{Felser2015}.
First, we consider an Hamiltonian containing the Hg d-states and anion p-states with the respective bare SOC defined $\lambda_{Hg-d}^{bare}$ and $\lambda_{a}^{bare}$ and we diagonalize the hamiltonian to calculate the effective spin-orbit $\lambda_{a}\propto$E($\Gamma_8$)-E($\Gamma_7$) as a function of the bare parameters.
We define the difference between the on-site energies ${\Delta}\epsilon_{1}$=$V_{pa}-V_{dc}>0$.
We calculate the eigenvalues of the Hamiltonian at the $\Gamma$ point.
In order to evaluate analytically the effective SOC, we perform the L\"owdin approximation projecting on the anion subspace obtaining:
\begin{equation}
\lambda_a\approx\lambda_a^{bare}-\frac{H_{p_xa,d_{yz}c}^2\lambda_{Hg-d}^{bare}}{({\Delta}\epsilon_{1}-\frac{3}{2}\lambda_{Hg-d}^{bare})({\Delta}\epsilon_{1}+\lambda_{Hg-d}^{bare})}
\end{equation}
where $H_{p_xa,d_{yz}c}$ is the matrix element that connects the $p_{x}-a$ and $d_{yz}-c$ orbitals calculated at the $\Gamma$ point.
If we consider the conditions
$|{\Delta}\epsilon_{1}| >>\lambda_{Hg-d}^{bare}$,
we obtain:
\begin{equation}\label{SOC4}
\lambda_a\approx\lambda_a^{bare}-\left(\frac{H_{p_xa,d_{yz}c}}{{\Delta}\epsilon_{1}}\right)^2\lambda_{Hg-d^{bare}}
\end{equation}
We can observe that the leading correction term to the bare spin-orbit is always negative.
The validity of this formula is restricted to the region close to the $\Gamma$ point, however, that is the interesting region for this class of compounds. 
This correction to the bare SOC is present in all the HgX (X=S,Se,Te) family, however, it is quantitatively more relevant in HgS.

\section{Analytical derivation of the end state energies and wave functions \label{app:end-states}}

In this section we calculate the eigenenergies and the wave functions for the end states of the minimal model Hamiltonian (\ref{eq:pbc}), when the system has a finite length $L$ along $k_1$ direction. By using the Fourier 
decomposition 
\begin{equation}
{\cal H}\left(k_{1},k_{3}\right)=H{}_{0}\left(k_{3}\right)+e^{ik_{1}}H_{1}\left(k_{3}\right)+e^{-ik_{1}}H_{1}^{\dagger}\left(k_{3}\right)
\end{equation}
the Hamiltonian for the system be written as 
\begin{equation}
{\cal H}^{1D}\left(k_{3}\right)\!=\!\mathbbm{1}_{L}\otimes H_{0}\left(k_{3}\right)\!+\!D\otimes H_{1}\left(k_{3}\right)\!+\!D^{\intercal}\otimes H_{1}^{\dagger}\left(k_{3}\right),
\end{equation}
where $D$ is $L\times L$ matrix of the form (\ref{eq:d}). The eigenstates  $|\psi \rangle = \sum_{j=1}^L |j \rangle \otimes |\phi_j\rangle$, where  $\left|j\right\rangle$ is the basis state describing the $j$th unit cell along the $k_1$ direction and $|\phi_j\rangle$ is the spinor describing the state within the unit cell, 
must satisfy 
\begin{equation}
H_{1}^{\dagger}\left(k_{3}\right)\left|\phi_{j-1}\right\rangle  +  H_{0}\left(k_{3}\right)\left|\phi_{j}\right\rangle +H_{1}\left(k_{3}\right)\left|\phi_{j+1}\right\rangle 
=E\left|\phi_{j}\right\rangle, \label{bulk-app-L}
\end{equation}
for $j=2,3,\dots,L-1$ and the boundary equations
\begin{eqnarray}
H_{0}\left(k_{3}\right)\left|\phi_{1}\right\rangle +H_{1}\left(k_{3}\right)\left|\phi_{2}\right\rangle  & = & E\left|\phi_{1}\right\rangle ,\nonumber \\
H_{1}^{\dagger}\left(k_{3}\right)\left|\phi_{L-1}\right\rangle +H_{0}\left(k_{3}\right)\left|\phi_{L}\right\rangle  & = & E \left|\phi_{L}\right\rangle.
\end{eqnarray}
We write an ansatz for the end states  ($n=1,2$) as 
\begin{eqnarray*}
\left|\phi_{n, j}\left( q\right)\right\rangle  & = & A_{q}e^{ijq}\left|E_{n}\left(q\right)\right\rangle +B_{q}e^{-ijq}\left|E_{n}\left(-q\right)\right\rangle   
\end{eqnarray*}
where   the energies  
\begin{equation}
E_{1(2)}\left(q\right)  =  \frac{1}{2}\left(\eta_{p}\!+\!\eta_{s}\! \mp \!\sqrt{56\!+\!\left(\eta_{p}\!-\!\eta_{s}\right)^{2}\!-\!8\cos q}\right).
\end{equation}
and spinors $|E_{1(2)}(q)\rangle$ are obtained from the dispersive bulk states by replacing the momentum $k_1$ with $q$.
Therefore, the ansatz automatically satisfies Eq.~(\ref{bulk-app-L}). Importantly, here $q$ is an
imaginary number so that the ansatz describes a state localized at the end of the system. 
Note that $\left|\phi_{n,j}\left(q\right)\right\rangle $ also contains
dependence on $k_{3}$ which is however not essential here. 
Using the boundary conditions we find in the thermodynamic limit $L\to \infty$ for  both $n=1,2$ and any $k_{3}$ that
$q=-i\ln 3$ so that 
the charge density has a decay length 
\begin{equation}
\xi=\frac{1}{2\log3}.
\end{equation}
All end-states localized at the right end of the
system, and their energies are 
\begin{equation}
E_{1(2)}\left( - i \ln 3\right) = \frac{1}{2}\left(\eta_{p}\!+\!\eta_{s}\! \mp \!\sqrt{\frac{128}{3}\!+\!\left(\eta_{p}\!-\!\eta_{s}\right)^{2}}\right).
\end{equation}

\section{$k_1$---$k_2$ duality of the minimal model \label{app:duality}}

Consider the minimal model obtained 
as in the Section \ref{sec:minimal}.
Now, instead of setting $k_2=\pi$ we
take $k_1=\pi$. The  2D bulk Hamiltonian of this system is 
\begin{equation}
{\cal \tilde{H}}^{2D}\left(k_{2},k_{3}\right)=\tilde{H}_{\parallel}+\left(e^{ik_{3}}\tilde{H}_{\perp}\left(k_{2}\right)+H.c.\right),
\end{equation}
and its spectrum does not depend on $k_3$ again. 
The 1D Hamiltonian of $W$ unit cells wide cylinder ($c=1$) is  
\begin{equation}
{\cal \tilde{H}}_{c=1}^{1D}\left(k_{2}\right)=\mathbbm{1}_{W}\otimes\tilde{H}_{\parallel}+\left(T\otimes\tilde{H}_{\perp}\left(k_{2}\right)+H.c.\right),
\label{eq:Htilde}
\end{equation}
which we will compare to the previous case of $k_2=\pi$ Hamiltonian
\begin{equation}
{\cal H}_{c=1}^{1D}\left(k_{1}\right)=\mathbbm{1}_{W}\otimes H_{\parallel}\left(k_{1}\right)+\left(T\otimes H_{\perp}+H.c.\right).
\label{eq:Hnotidle}
\end{equation}
Here $T$ is defined by Eq.~(\ref{eq:Tpbc}),  $H_{\parallel}(k_1)$ and $H_{\perp}$ are defined in Section \ref{sec:minimal}, and 
\begin{equation}
    \frac{\tilde{H}_{\parallel}}{E_{0}}=\begin{pmatrix}\eta_{s} & -2 & 2 & 0\\
-2 & \eta_{p} & 0 & 0\\
2 & 0 & \eta_{p} & 0\\
0 & 0 & 0 & \eta_{p}
\end{pmatrix},
\end{equation}
and
\begin{equation}
    \frac{\tilde{H}_{\perp}\left(k_{2}\right)}{E_{0}}=\begin{pmatrix}0 & -e^{-ik_{2}}\!+\!1 & -e^{-ik_{2}}\!+\!1 & e^{-ik_{2}}\!+\!1\\
0 & 0 & 0 & 0\\
0 & 0 & 0 & 0\\
0 & 0 & 0 & 0
\end{pmatrix}.
\end{equation}
By a close inspection of the 1D chains
described by the Hamiltonians (\ref{eq:Htilde},\ref{eq:Hnotidle}) we find a duality relation between these two Hamiltonian given by 
\begin{equation}
    {\cal H}_{c=1}^{1D}\left(k_{1}\right)={\cal V}^{\dagger}{\cal \tilde{H}}_{c=1}^{1D}\left(k_{2}=-k_{1}\right){\cal V},
    \label{eq:dual}
\end{equation}
where $\cal V$ is a unitary operator
\begin{equation}
{\cal V}=\begin{cases}
P_{3}P_{2}P_{1}GQ & W\in2\mathbb{N}+1\\
P_{3}P_{2}P_{1}G & W\in2\mathbb{N}
\end{cases}.
\end{equation}
Here $P_{1,2,3}$ are the site permutations 
\begin{equation}
P_{1}=\mathbbm{1}_{W}\otimes\begin{pmatrix}0 & 1 & 0 & 0\\
0 & 0 & 1 & 0\\
0 & 0 & 0 & 1\\
1 & 0 & 0 & 0
\end{pmatrix},
\end{equation}
\begin{equation}
P_{2}=\begin{pmatrix}0 & \cdots & 0 & 0 & 0 & 1 & 0\\
0 & \cdots & 0 & 0 & 1 & 0 & 0\\
0 & \cdots & 0 & 1 & 0 & 0 & 0\\
\vdots & \vdots & \vdots & \vdots & \vdots & \vdots & \vdots\\
0 & 1 & 0 & 0 & \cdots & 0 & 0\\
1 & 0 & 0 & 0 & \cdots & 0 & 0\\
0 & 0 & 0 & 0 & \cdots & 0 & 1
\end{pmatrix}
\end{equation}
and
\begin{equation}
    P_{3}=\mathbbm{1}_{W}\otimes\begin{pmatrix}0 & 0 & 0 & 1\\
0 & 1 & 0 & 0\\
0 & 0 & 1 & 0\\
1 & 0 & 0 & 0
\end{pmatrix}.
\end{equation}
Note that $P_1$ and $P_3$ only reshuffles sites inside the unit cell consisting of four sites whereas the order of the unit cells remains unchanged. On the other hand, the $P_2$  operator reverts the whole chain and shifts sites by one in a cyclic manner.
The remaining constituents of $\cal V$ are the alternating gauge matrix
\begin{equation}
    G=\begin{pmatrix}1 & 0 & 0 & 0 & \cdots\\
0 & -1 & 0 & 0\\
0 & 0 & 1 & 0\\
0 & 0 & 0 & -1\\
\vdots &  &  &  & \ddots
\end{pmatrix}\otimes\begin{pmatrix}1 & 0 & 0 & 0\\
0 & -1 & 0 & 0\\
0 & 0 & 1 & 0\\
0 & 0 & 0 & -1
\end{pmatrix},
\end{equation}
and operator $Q$ acting on the first unit cell 
\begin{eqnarray}
Q & = & \begin{pmatrix}1 & 0 & 0 & \cdots & 0\\
0 & 0 & 0 & \cdots & 0\\
0 & 0 & 0 & \cdots & 0\\
\vdots & \vdots & \vdots & \vdots & \vdots\\
0 & 0 & 0 & \cdots & 0
\end{pmatrix}\otimes\begin{pmatrix}1 & 0 & 0 & 0\\
0 & 0 & -1 & 0\\
0 & -1 & 0 & 0\\
0 & 0 & 0 & 1
\end{pmatrix}\nonumber \\ 
 & + & \begin{pmatrix}0 & 0 & 0 & 0 & \cdots & 0\\
0 & 1 & 0 & 0 & \cdots & 0\\
0 & 0 & 1 & 0 & \cdots & 0\\
\vdots & \vdots & \vdots & \vdots & \vdots & \vdots\\
0 & 0 & \cdots & 0 & 1 & 0\\
0 & 0 & 0 & \cdots & 0 & 1
\end{pmatrix}\otimes\mathbbm{1}_{4}.
\end{eqnarray}
From Eq.~(\ref{eq:dual}) and results of Section \ref{sec:minimal} it follows that for a cylindrical geometry along $k_3$ the ${\cal \tilde{H}}_{c=1}^{1D}\left(k_{2}\right)$ Hamiltonian splits into 
$W$ identical $1\times1$ blocks 
\begin{equation}
\tilde{B}_1=E_0 (\eta_p),
\end{equation}
and another $W$ identical $3 \times 3$ blocks  
\begin{eqnarray}
\tilde{B}_{3} & = & E_0 \begin{pmatrix}\eta_{p} & \sqrt{6-2\cos k_2} & 0\\
\sqrt{6-2\cos k_2} & \eta_{s} & \sqrt{8}\\
0 & \sqrt{8} & \eta_{p}
\end{pmatrix},
\end{eqnarray}
under transformation $\cal W = V\,U$.
The same transformation $\cal W$ used on a ribbon-geometry ($c=0$) Hamiltonian  
\begin{equation}
{\cal \tilde{H}}_{c=0}^{1D}\left(k_{2}\right)=\mathbbm{1}_{W}\otimes\tilde{H}_{\parallel}+\left(D\otimes\tilde{H}_{\perp}\left(k_{2}\right)+H.c.\right)
\label{eq:Htidle}
\end{equation}
with $D$ defined by Eq. (\ref{eq:d})
gives almost the same block structure with one of the $3 \times 3$ blocks substituted by
\begin{eqnarray}
\tilde{B}^{\prime}_{3} & = & E_0 \begin{pmatrix}\eta_{p} & 0 & 0\\
0 & \eta_{s} & \sqrt{8}\\
0 & \sqrt{8} & \eta_{p}
\end{pmatrix}.
\end{eqnarray}
Concluding, we have found that for cylindrical geometry the Hamiltonians 
${\cal H}_{c=1}^{1D}\left(k_{1}\right)$ and ${\cal \tilde{H}}_{c=1}^{1D}\left(k_{2}\right)$  have the same band structures and the same end-states. In the case of the ribbon geometry  ${\cal \tilde{H}}_{c=0}^{1D}\left(k_{2}\right)$ has two more (less) flat (dispersive) bands compared to ${\cal H}_{c=0}^{1D}\left(k_{1}\right)$, following from the difference between blocks $\tilde{B}^{\prime}_{3}$ and $B^{\prime}_{3}$. The number of end-states in this case is smaller by two because the $\tilde{B}^{\prime}_{3}$ having no dispersion cannot contribute any, unlike $B^{\prime}_{3}$.

\bibliography{QW}

\end{document}